\documentclass[conference]{IEEEtran}
\usepackage{amssymb,amsmath,epsfig}
\usepackage{graphicx,epstopdf,subfigure}
\usepackage{citesort}
\usepackage{times}

\newtheorem{theorem}{Theorem}

\newtheorem{problem}{Problem}

\newcommand{\mcal}[1]{{\mathcal{#1}}}
\newcommand{\mbb}[1]{{\mathbb{#1}}}
\newcommand{\mbf}[1]{{\mathbf{#1}}}

\newcommand{\field}[1]{\mbb{#1}}

\renewcommand{\r}{{\mathbf{r}}}
\renewcommand{\k}{{\mbf{k}}}
\renewcommand{\j}{{\iota}}

\newcommand{\ceil}[1]{{\left\lceil#1\right\rceil}}

\newcommand{\ie}{{\emph{i.e.}}}

\graphicspath{{.}{./figs/}}
\centerfigcaptionstrue

\newcommand{\footnotethanks}[1]{%
\renewcommand{\thefootnote}{\empty}%
\footnotetext{#1}%
\renewcommand{\thefootnote}{\arabic{footnote}}
\setcounter{footnote}{1}}%

\begin{document}

\title{Bounds on Space-Time-Frequency Dimensionality}
%
\author{
	\authorblockN{Leif W. Hanlen$^\dag$ and Thushara D. Abhayapala$^\dag$}%
	\authorblockA{National ICT Australia and Australian National University\\
	{\tt \{Leif.Hanlen,Thushara.Abhayapala\}@nicta.com.au}}
}
 
\maketitle

\begin{abstract}
We bound the number of electromagnetic signals which may be observed over a frequency range $2W$ for a time $T$ within a region of space enclosed by a radius $R$. Our result implies that broadband fields in space cannot be arbitrarily complex: there is a finite amount of information which may be extracted from a region of space via electromagnetic radiation.

Three-dimensional space allows a trade-off between large carrier frequency and bandwidth.  We demonstrate  applications in super-resolution and broadband communication.
\end{abstract}

\section{Introduction}

\footnotethanks{${}^\dag$ L. W. Hanlen and T. D. Abhayapala also hold appointments with the Research School of Information Sciences and Engineering, ANU. 
 %
National ICT Australia is funded through the Australian Government's \emph{Backing Australia's Ability initiative}, in part through the
Australian Research Council.}

The concept of ``rich multipath'' as a positive communication medium is widely established in antenna~\cite{AndrewsMitradeCarvalho02,Bucci:12:1987,Bucci:07:1989} signal processing~\cite{JonKenAbh:ICASSP02,SadeAbhaKenn:ICASSP2006} and communication~\cite{Moustakas00,Sayeed02,PoonBroTse2005} literature. It is well known that ``rich'' multipath allows multiplexing of independent signals. 
 %
Dimensionality has been offered as an effective measure of richness in the narrow-band regime~\cite{JonKenAbh:ICASSP02,HanlTimo:ITW2006}. The results, however, have required a (very) narrow bandwidth signal --  the signal is defined by its wavelength.

The spatial bounds of~\cite{JonKenAbh:ICASSP02,SadeAbhaKenn:ICASSP2006} were  based on the observation~\cite{Colton98,ColtKress1995} that (narrowband) far-field radiative waves are functions which may be bounded (exponentially) toward zero beyond some limit. Broadband representations of signals~\cite{Bucci:03:1998} have been developed,  these works are not conducive to analytic dimensionality bounds.
 
For non-spatially diverse signals, dimensionality is well developed. Shannon~\cite{Shannon1949}  conjectured that a signal with bandwidth $2W$ which was observed for a time $T$ could be represented by $2WT$ parameters. This provided the concept of a time-bandwidth product and the $2WT+1$ result was formalised by several authors~\cite{Dollard64,Pollak0161,Slepian64,Slep:1976,bennett:1969}. 
 %
There are an infinity of signals which can be placed in a bandwidth $2W$, and likewise an infinity which may be separately observed in a time interval $T$. However only a discrete set of signals may be resolved in a $2W$ bandwidth within a time $T$.

Dimensionality results (and subsequent capacities~\cite{WalJen02VTCb,HanlTimo:ITW2006}) for single-frequency spatial communication provide an intuition of the fundamental limits to spatial multiplexing. However practical spatial signalling techniques -- such as MIMO with Orthogonal Frequency Division Multiplexing (OFDM) or Ultra-Wide Band (UWB) -- motivate  the dimensionality of spatial signals with significant bandwidth components. 

\begin{quote}
\emph{What is the limit to the number of (radio) signals which may be detected in a region of space, when the only constraints are the region size, signal bandwidth and time of observation?} 
\end{quote}

 %

Determining the number of electromagnetic signals which may be detected within a region of space allows development of practical limits to multiple-antenna communication. This question also addresses several open problems: 
\begin{enumerate}
\item Broadband spatial communication (eg. MIMO-OFDM and MIMO-UWB): how can/should one trade space for frequency/time? 

\item Broadband beamforming~\cite{UthaBial:AP:022006} and so-called ``super-resolution'' results~\cite{Lukosz66}: can one resolve at better than the spatial (optical) resolution limit of~\cite{JonKenAbh:ICASSP02,Miller0400} and if so, how (and why)?

\item Small space: Is there value in using the 3D spatial dimension, when the spatial extent is much less than the signal wavelength \ie\ $R \ll \lambda_{\min}$?

\end{enumerate}

For the purpose of this paper we assume the ``antenna'' occupies the entire spatial region of interest, and that filters and truncations are ideal. We neglect mutual coupling.

The remainder of this paper is arranged as follows: Section~\ref{s:background} provides an introduction to the heuristic approach, in the context of the well-known $2WT+1$ dimensionality result. Section~\ref{s:dimension} provides a dimensionality result for 3D- and 2D-space. We provide discussion in Section~\ref{s:discuss} and draw conclusions in Section~\ref{s:conclusion}. Proofs are contained in the appendix.


%



\section{ Background} \label{s:background}
 
Current dimensionality results for spatially and time-bandwidth constrained single-frequency signals are:
\begin{align}
\mcal{D}_{\text{2D space}} &= \left\lceil\frac{e\pi F R}{c}\right\rceil+1 \label{E:Kennedy2d}
\\
\mcal{D}_{\text{3D space}} &= \left(\left\lceil\frac{e\pi F R}{c}\right\rceil+1\right)^2 \label{E:Kennedy3d}
\\
\mcal{D}_{2WT} &= \left\lceil2WT\right\rceil+1\label{E:2wt}
\end{align}
where \eqref{E:Kennedy2d} and \eqref{E:Kennedy3d} are derived in~\cite{JonKenAbh:ICASSP02,Kennedy0202}, although \eqref{E:Kennedy3d} is corrected from~\cite{Kennedy0202}. 
 %
It is incorrect to simply multiply (say) \eqref{E:Kennedy3d} and \eqref{E:2wt} to find the dimensionality of spatial signals with a finite bandwidth. Equally, integrating  a spatial dimensionality result~\eqref{E:Kennedy2d} over the range $F_o-W\leq f\leq F_o+W$ leads to the incorrect result of a ``dimensionality'' which is not unit-less.
 
We recall a heuristic argument~\cite{Gallager68,Shannon1949}  for the number of orthogonal signals which may be confined into a $2W\times T$ time-bandwidth product.
 
 \smallskip
\begin{problem}[Shannon $2WT$]\label{Prob:2wt}
\emph{Find the number $\mcal{D}_{2WT}$ of orthogonal signals which may be observed within a bandwidth $F_o\pm W$ over a time interval $T$.}
\end{problem}
\medskip

\begin{figure}
\centering 
\setlength{\unitlength}{0.50mm}
\begin{picture}(110,70)
\put(0,0){\includegraphics[width=55mm]{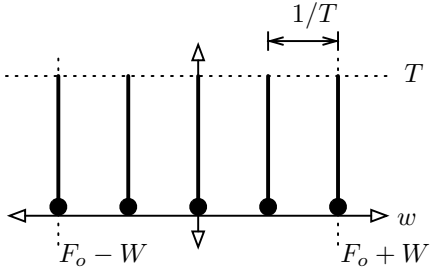}}
\put(77,64){$1/T$}
\put(107,48){$T$}
\put(105,10){{$w$}}
\put(15,0){$F_o-W$}
\put(90,0){$F_o+W$}
\end{picture}
\caption{Time-Frequency block, $-W\leq f\leq W$, $0\leq t\leq T$. functions are continuous in time, but chosen at discrete intervals in frequency. Dimensionality is given by counting the solid circles.}\label{F:wt}
\end{figure}

Figure~\ref{F:wt} describes the solution to the problem graphically: a signal which is constrained to $2W\times T$ comprises continuous time functions, which occur at a spacing of $1/T$ from lower frequency $F_o-W$ to upper frequency $F_o+W$. Counting the number of functions gives the dimension of the original signal. The proof is not developed by counting $2W\times T$ ``small'' $1\mathrm{Hz}\times1\mathrm{s}$ blocks: it is given by counting the number of functions (lines) in Figure~\ref{F:wt} which are separated by $1/T$.

Following~\cite[Chap.8]{Gallager68}, write an arbitrary signal $x(t)$, \mbox{$t\in[0,T]$} as:
\begin{align}
x(t) &= \sum_{m=-\infty}^{\infty} \alpha_m \phi_m(t) \label{E:DFT}
\\ %
\phi_m(t)&=
\begin{cases}
\frac{1}{\sqrt{T}}\exp\left(\j \frac{2\pi m t}{T}\right) &;\ t\in[0,T]\\
0 &;\ \text{else}
\end{cases}
\end{align}
for integer $m\in\field{Z}$.
 %
Observe $\phi_m(t)$ is a time-limited sinusoid, with frequency $m/T$, and \eqref{E:DFT} defines a discrete Fourier transform. In order to limit $x(t)$ to having frequencies in the range $F_o-W\cdots F_o+W$ we restrict $\alpha_m$ to be non-zero for only appropriate values for $m$:
\begin{equation}
F_o-W< \frac{m}{T} < F_o+W
\end{equation}
which implies:
\begin{equation}\label{E:xtsum}
x(t) =\sum_{m=(F_o-W)T}^{(F_o+W)T} \alpha_m \phi_m(t)
\end{equation}
There are \emph{at most} $$(F_o+W)T-(F_o-W)T+1 = 2WT+1$$ non-zero terms in~\eqref{E:xtsum}, which implies $\mcal{D}_{2WT}=\lceil 2WT \rceil+1$

This heuristic argument is not precise since the function $x(t)$ may not be well behaved in terms of finite- bandwidth and time constraints: it is possible to choose a particular $x(t)$ such that more than $2WT+1$ sinusoid functions $\phi_m(t)$ are required to parameterise $x(t)$, since the Fourier series are not complete on the interval $[0,T]$. However, in the limit of large $W$ and large $T$, the $2WT$ approximation becomes accurate~\cite{Slep:1976}. The effect of ``badly behaved'' $x(t)$ is detailed in~\cite{Pollak0161,Pollak0161b,Pollak0162}

\section{Dimensionality: $R\times2W\times T$}\label{s:dimension}

Consider an arbitrary electromagnetic field observed within a fixed region of space, of radius $R$, for a time $T$ and over a frequency  range $F_o\pm W$, where $F_o$ is the centre frequency. We wish to calculate the number $\mcal{D}$ of orthogonal signals which may be observed. 
In addition, the signals must be propagating electric fields. 
 
Source-free (propagating) electric fields $\psi(\r,t)$, satisfy a reduced form of the wave equation, known as the Helmholtz wave equation~\cite{Colton98,Gradshteyn00}:
\begin{equation}\label{E:helmholtz}
\left(\nabla^2 - \frac{1}{c^2} \frac{\partial^2}{\partial t^2}\right)\cdot\psi(\r,t)=0
\end{equation}
where $\nabla=(\partial/\partial x, \partial/\partial y, \partial/\partial z)$ is the gradient and \mbox{$c=3\times10^8$} is the speed of light. 

 %
\begin{problem}
\begin{itshape}
Given a function in space-time $x(\r,t)$ which is non-zero for $\|\r\|\leq R$ and $t\in[0,T]$  has a frequency component in $[F_o-W,F_o+W]$ and satisfies \eqref{E:helmholtz}; what number $\mcal{D}$ of orthogonal signals $\varphi(\r,t)$ are required to parameterize $x(\r,t)$?

\ie\ \emph{what is the dimension $\mcal{D}$ of the space of wave-signals constrained by $R\times2W\times T$?}
\end{itshape}
\end{problem}


\medskip
\begin{theorem}[3D dimensionality $\mcal{D}_{3D}$]\label{thm:2}
The number of orthogonal electromagnetic waves which may be observed in a three-dimensional spatial region  bounded by radius $R$, over frequency range $F_o\pm W$ and time interval $[0,T]$ is
\begin{multline}\label{E:thm2}
\mcal{D}_{3D}\leq 
 TW\left[
\left(\frac{2W^2}{3}
+2 F_o^2 \right)\left(\frac{e\pi R}{c}\right)^2
+\frac{6e\pi R F_o}{c}
+\frac{13}{3}\right]
\\ + \left( \frac{(F_o-W) e \pi R}{c}+1\right)^2
\end{multline}
\end{theorem}
\medskip

\subsection{Asymptotics}

For $TW\to0$, Theorem~\ref{thm:2} reduces to \eqref{E:Kennedy3d}, while for \mbox{$R\to0$,} Theorem~\ref{thm:2} reduces to $13TW/3+1$ which over-bounds the $2WT+1$ result.
Asymptotically ($T,W,R\to\infty$), Theorem~\ref{thm:2} may be approximated by:
\begin{equation}\label{E:3d-asymptote}
\mcal{D}_{3D}\to2TW
\left(\frac{W^2}{3}
+ F_o^2 \right)\left(\frac{e\pi R}{c}\right)^2
\end{equation}

In Problem~\ref{Prob:2wt}, the carrier frequency is irrelevant: since the whole problem may be pre-modulated by the scalar function $\exp(-\j 2\pi F_o t)$, without altering the  degrees of freedom. In the case of the wave equation, time and space are linked: altering the carrier frequency alters the degrees of freedom, independent of the bandwidth. This can be seen in \eqref{E:3d-asymptote} where   both centre-frequency and bandwidth may be traded against each other.

\subsection{2D space}

\smallskip
\begin{theorem}[2D dimensionality $\mcal{D}_{2D}$]\label{thm:1}
The number of orthogonal electromagnetic waves which may be observed in a two-dimensional  spatial region  bounded by radius $R$, over frequency range $F_o\pm W$ and time interval $[0,T]$ is
\begin{equation}
\mcal{D}_{2D} \approx \underbrace{4WT}_{\text{time-band}}+1 +\underbrace{\frac{2 e \pi R}{c}   F_o}_{\text{narrowband space}} + \underbrace{\frac{2 e \pi R}{c} TW^2}_{\text{broadband space}}
\end{equation}
\end{theorem}
\medskip


\section{Discussion}\label{s:discuss}

\begin{figure}
\centering 
\subfigure[%
Space-Frequency block, $-W\leq f\leq W$, $0\leq r\leq R$.  \label{F:rw}
]{
\setlength{\unitlength}{0.60mm}
\begin{picture}(125,60)
\put(0,0){\includegraphics[width=60mm]{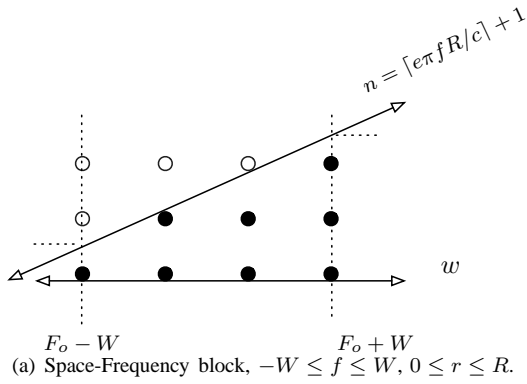}}
\put(100,15){{$w$}}
\put(77,-2){\footnotesize $F_o+W$}
\put(12,-2){\footnotesize $F_o-W$}
\put(82,55){\rotatebox{25}{\footnotesize $n=\ceil{e\pi f R/c}+1$}}
\end{picture}
}

\subfigure[%
Space-Time-Frequency block. Note that the block is a trapzeoid: more spatial functions appear for higher frequencies. Figure~\ref{F:rw} shows the front trapezium face of the block.\label{F:rtw}
]{%
\setlength{\unitlength}{0.45mm}
\begin{picture}(155,110)
\put(10,0){\includegraphics[width=45mm]{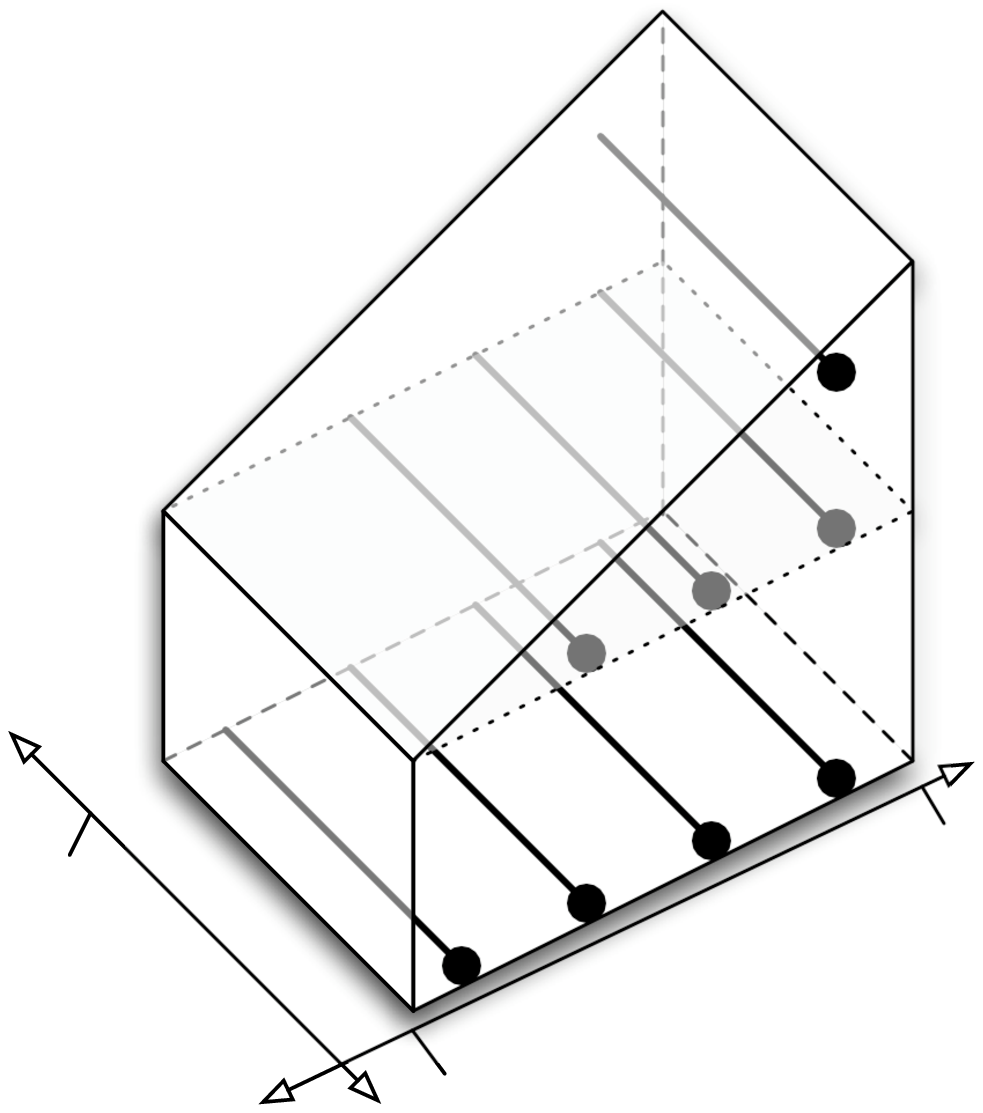}}
\put(102,22){\footnotesize $F_o+W$}
\put(57,2){\footnotesize $F_o-W$}
\put(17,22){\footnotesize $T$}
%
\end{picture}
}
\caption{Space-Time-Frequency. Functions are continuous in time, and chosen discretely over frequency and space, via the indexing process below. Dimensionality given by counting solid circles. The arrangement of functions in Figure~\ref{F:wt} form  \emph{horizontal layers} $\r=\mathrm{constant}$ in the trapezoid.}\label{F:spacetimefreq}
\end{figure}

In both the 2D and 3D cases, we use Figure \ref{F:spacetimefreq} as guide, in the same manner as Figure~\ref{F:wt} is used for the $2WT$ result. In the case of the $2WT$, the constraints formed a block, side-length $2W\times T$. Here, we consider a three-dimensional set of constraints, over frequency $F_o\pm W$, space $R$ and time $T$. The functions (lines) are continuous in time, and chosen discretely over space and frequency. Considering the space-frequency face Figure~\ref{F:rw}, the constraints are given by $F_o-W\leq f\leq F_o+W$, and the number of spatial functions is given by $n\leq(e\pi fR)/c$. This line forms the top edge of a trapezium -- since the number of functions which can be observed in space increases as the centre frequency increases. 

Theorem~\ref{thm:2} may be interpretted as a packing problem on the surface of a sphere. For non-trivial time, space and bandwidth constraints the result \eqref{E:thm2} is dominated by the first two terms,  
\begin{equation}\label{E:halfa}
2WTF_o^2 \left(\frac{e\pi R}{c}\right)^2
\end{equation}
 and 
\begin{equation}\label{E:halfb}
\left(\frac{2}{3}TW^3\right)\left(\frac{e\pi R}{c}\right)^2.
\end{equation}
 
The   term  \eqref{E:halfa} is very intuitive: the dimension of spatial signals, of a single frequency $f=ck/2\pi$ is proportional to the surface area of the region in wavelengths~\cite{JonKenAbh:ICASSP02} and the dimension of time-frequency signals, constrained to a point~\cite{Gallager68,Pollak0161} is $2WT$. Equation \eqref{E:halfa}  can be seen as the multiplication of \eqref{E:Kennedy3d} with \eqref{E:2wt}. We note that it is the surface-area of the region not the volume which defines the dimensionality.

For the  component \eqref{E:halfb}, consider launching  wavelike photons from the centre of $\mbb{S}$: each has an outward velocity $c$, and contains $WT$ independent signals.
Interpret $R/c = \tau$ as the \emph{time in seconds} before each photon will cross the boundary of $\mbb{S}$ (and thus no longer be observed). Each photon may be viewed as a meta-signal, with time-bandwidth $W\tau$. As the surface of $\mbb{S}$ is 2 dimensional, we have 2 orthogonal sets of these meta-signals -- effectively separated by space. So we may observe $(W\tau)\times(W\tau)=(W\tau)^2$ independent photons, each containing $WT$ independent signals.  Giving $(\tau W)^2 \times (TW)$ independent signals. 


\subsection{Asymptote in 3D}
The average number of spatial modes  per Hz in 3D between $F_o-W$ and $F_o+W$ is 
\begin{equation}\label{E:av.mode}
\left(F_o^2+W^2\right)\left(\frac{e\pi R}{c}\right)^2
\end{equation}
For large $T,R,W$ the constraints becomes less onerous, and (as in the case for time-bandwidth) all three dimensions (space-time-frequency) become essentially independent. In this case, the degrees of freedom is (essentially) the average number of spatial modes, multiplied by a constant $2WT$ degrees of freedom per time-band limited signal.
If we multiply \eqref{E:av.mode} by $2WT$ we have (approximately) the result of \eqref{E:3d-asymptote}.

\subsection{Plots}
\begin{figure}
\begin{center}
\setlength{\unitlength}{0.0040\textwidth}
\subfigure[DoF vs bandwidth $W$ and region radius $R$]{%
\begin{picture}(110,80)
\put(0,0){\includegraphics[width=0.40\textwidth]{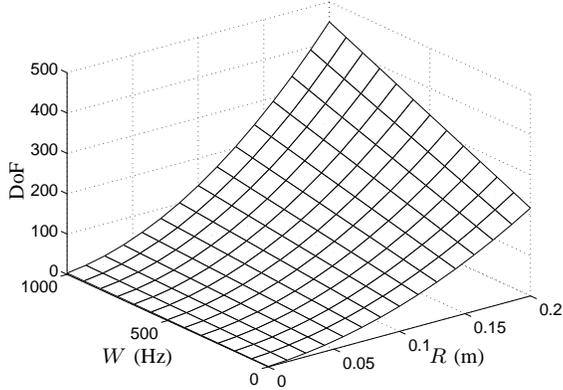}}
\put(75,5){\footnotesize{$R$ (m)}}
\put(15,5){\footnotesize{$W$ (Hz)}}
\put(-2,35){\rotatebox{90}{\footnotesize{DoF}}}
\end{picture}
}
\subfigure[Contours\label{F:largeWb}]{ %
\begin{picture}(110,80)
\put(0,3){\includegraphics[width=0.40\textwidth]{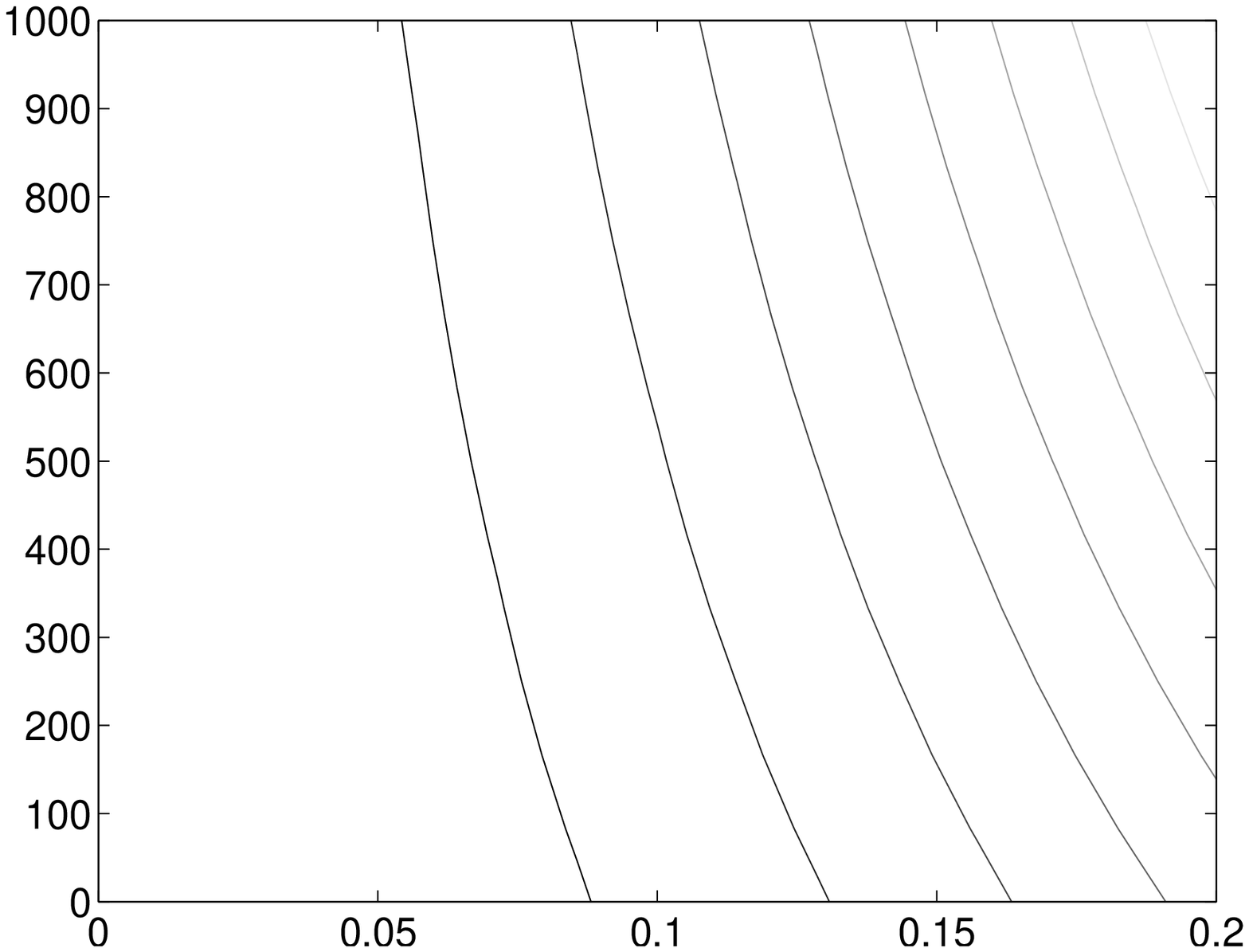}}
\put(45,-2){\footnotesize{$R$ (m)}}
\put(-3,35){\rotatebox{90}{\footnotesize{$W$ (Hz)}}}
\end{picture}
}
\caption{Number of degrees of freedom for moderate $W$ and $T=0.5ms$, $F_o=2.4GHz$, $\lambda_o=0.125m$: Curvature toward bottom of Fig.~\ref{F:largeWb} denotes saturation wrt. radius.}\label{F:largeW}
\end{center}
\end{figure}

\begin{figure}
\begin{center}
\setlength{\unitlength}{0.0040\textwidth}
\subfigure[DoF vs bandwidth $W$ and region radius $R$]{%
\begin{picture}(110,80)
\put(0,0){\includegraphics[width=0.40\textwidth]{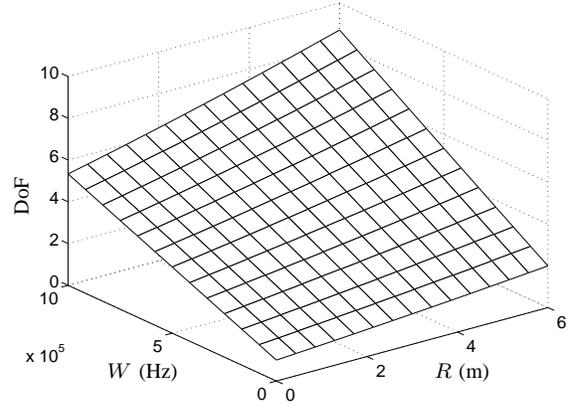}}
\put(75,5){\footnotesize{$R$ (m)}}
\put(15,5){\footnotesize{$W$ (Hz)}}
\put(-2,35){\rotatebox{90}{\footnotesize{DoF}}}
\end{picture}
}
\subfigure[Contours\label{F:smallWb}]{%
\begin{picture}(110,80)
\put(0,3){\includegraphics[width=0.40\textwidth]{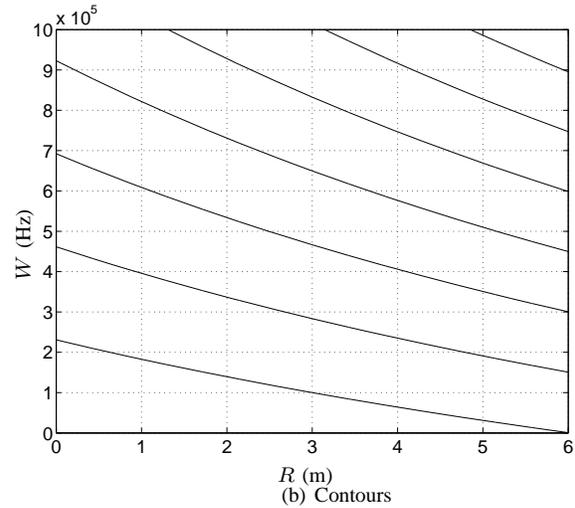}}
\put(45,-2){\footnotesize{$R$ (m)}}
\put(-3,35){\rotatebox{90}{\footnotesize{$W$ (Hz)}}}
\end{picture}
}
\caption{Number of degrees of freedom for large $W$, small $R$ and $T=1\mu s$, $F_o=2.4MHz$, $\lambda_o=125m$: Increase is almost linear in $R$ and $W$. }\label{F:smallW}
\end{center}
\end{figure}

\begin{figure}
\begin{center}
\setlength{\unitlength}{0.0040\textwidth}
\subfigure[Contours]{%
\begin{picture}(100,85)
\put(0,2){\includegraphics[width=0.40\textwidth]{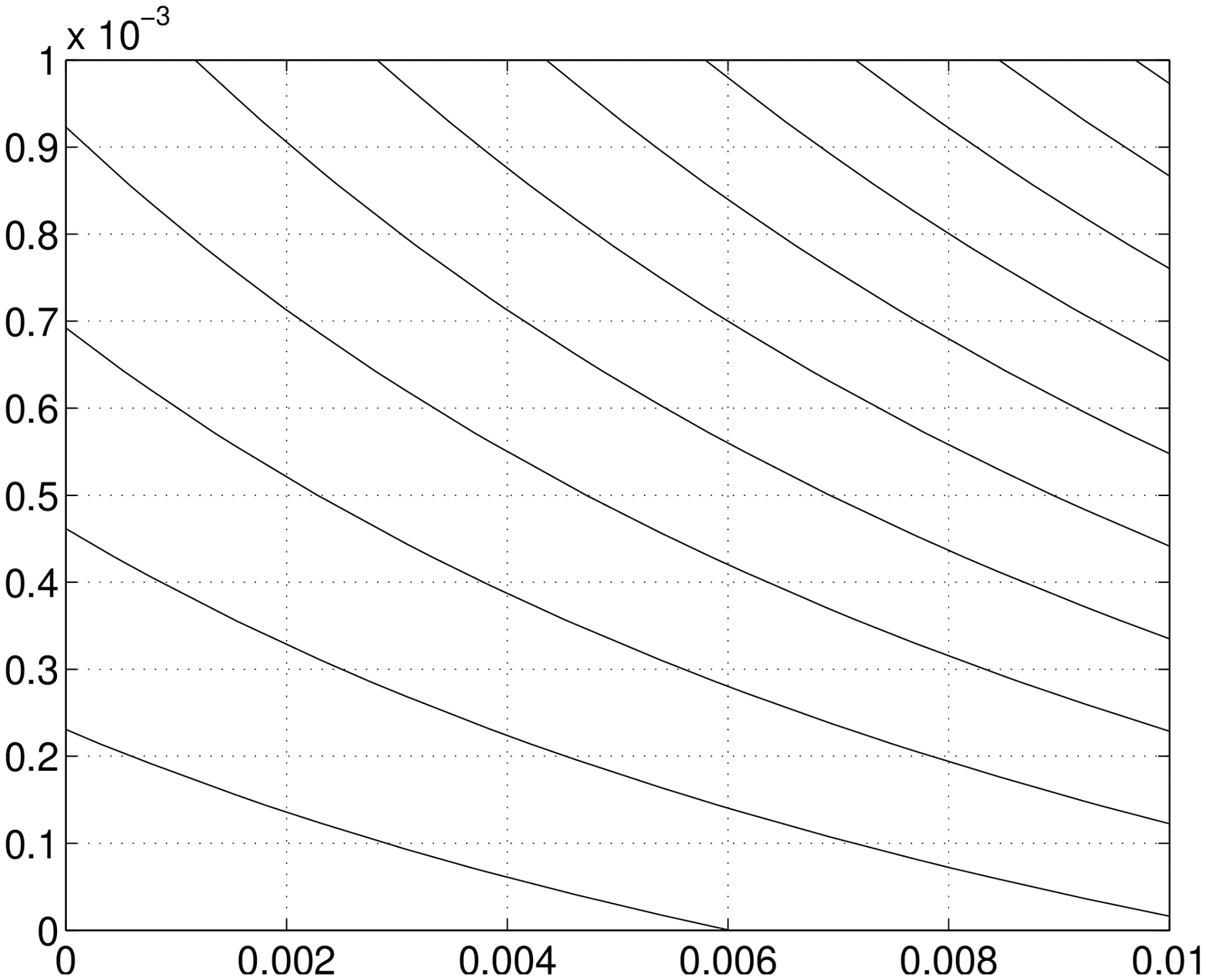}}
\put(45,-4){\footnotesize{$R$ (m)}}
\put(-4,35){\rotatebox{90}{\footnotesize{$T$ (s)}}}
\end{picture}
}
\subfigure[Degrees of Freedom vs Region Size $R$, for various observation time lengths $T$. At $T=0$ we have the (classical) result of \cite{Kennedy0202}.]{%
\begin{picture}(100,75)
\put(0,2){\includegraphics[width=0.40\textwidth]{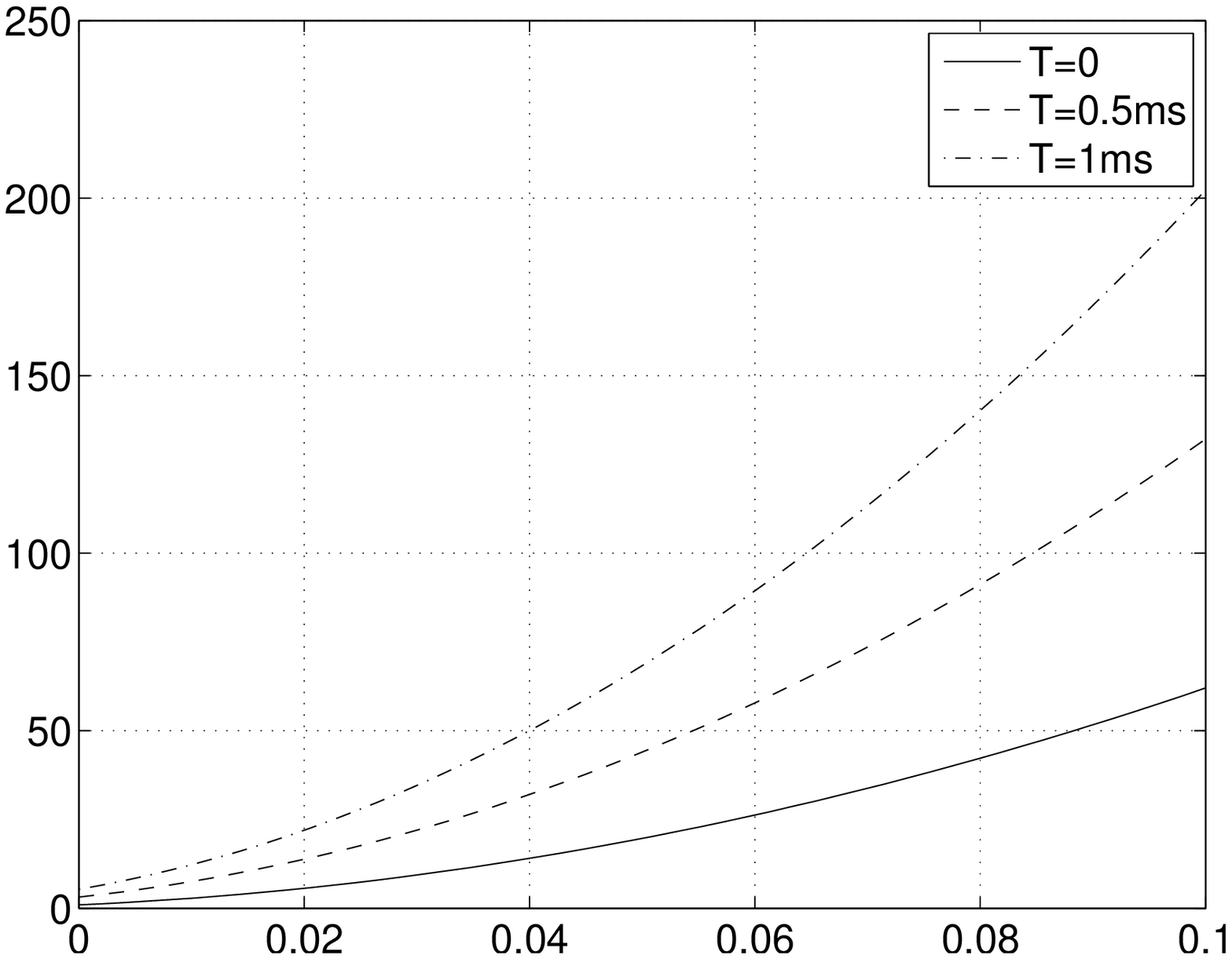}}
\put(45,-4){\footnotesize $R$ (m)}
\put(-4,35){\rotatebox{90}{\footnotesize DoF}}
\end{picture}
}
\subfigure[Degrees of Freedom vs Observation Time $T$, for various region sizes $R$. At $R=0$ we have the $2WT$ result.]{%
\begin{picture}(100,85)
\put(0,2){\includegraphics[width=0.40\textwidth]{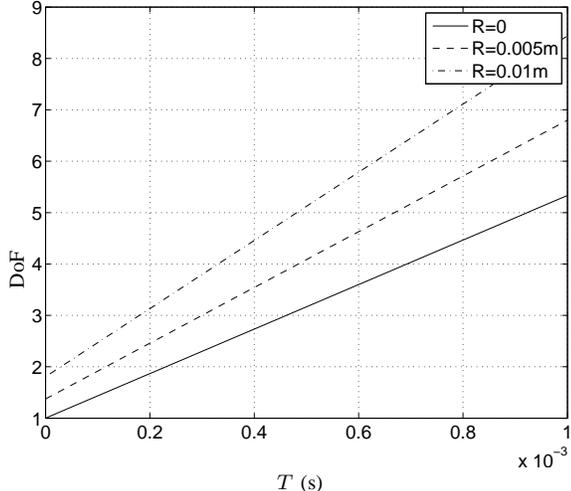}}
\put(45,-2){\footnotesize $T$ (s)}
\put(-4,35){\rotatebox{90}{\footnotesize DoF}}
\end{picture}
}
\caption{Number of Degrees of Freedom (DoF) vs observation time $T$ and region size $R$.
$W=$ 1kHz and $F_o=$ 2.4GHz.}\label{F:tr}
\end{center}
\end{figure}

We have given plots of the form of \eqref{E:thm2} in Figures~\ref{F:largeW}, \ref{F:smallW} and \ref{F:tr}. In Figure~\ref{F:largeW} we have considered a fixed carrier frequency $F_o=2.4$ GHz  fixed observation time $T=1\mu s$, and plotted the degrees of freedom for increasing region size $R$ and signal bandwidth $W$. It can be seen that the DoF increases dramatically once $R$ is beyond a single wavelength, moreover, moderate bandwidth signals give a super-linear increase in DoF. 

In Figure~\ref{F:smallW} we have set $F_o=2.4MHz$, with centre wavelength $\lambda_o=125m$ (thus causing $R$ to be much less than a wavelength). In this case, the super-linear terms do not eventuate, and DoF is approximately $2WT + \mcal{D}_{\text{3D space}}$. The (slight) benefit of increasing $R$ can be seen by the angle of the contours (not horizontal). The DoF is approximately linear in both $R$ and $W$.

For Figure~\ref{F:tr} we have considered the application of super-resolution or broad-band beamforming: by observing a signal for a sufficiently long time, we can extract more information than would be available by using the (classical) optical limit. In this case, space acts as a pre-multiplier, or gain, for what is essentially a time-bandwidth problem.

%

\section{Conclusion}\label{s:conclusion}

We have given preliminary bounds on the number of degrees of freedom (DoF) for broadband, spatially constrained signals observed over a finite time interval. Our results have been provided for both 2D and 3D spatial scenarios. For asymptotically large constraints (large volumes, large bandwidth and large time intervals) the DoF is given by a multiplication of standard DoF results for time and space. For small constraints, the DoF result collapses to a sum of DoF's for space and time.


\appendix

\begin{proof}[Theorem~\ref{thm:2}]

\begin{itshape}
\begin{enumerate}
\item Write a time-space function $x(\r,t)$ in terms of functions $$\psi_i(\r,t) = \varphi_i\bigl(\r;k(i)\bigr) \exp\left[-\j  k(i) c t\right]$$ where $k(i)$ is the scalar wave-number, and $i$ is an index.
\item Observe that every spatial function $\Psi_i(\r,t)$ which satisfies \eqref{E:helmholtz} in three dimensions has $$\mcal{D}_{\text{3D Space}}(i) = \left(\frac{e \pi F(i) R}{c} + 1\right)^2$$ degrees of freedom, with frequency $F(i)$, radius $R$.

\item The total degrees of freedom is  (equivalent to \eqref{E:xtsum})
$$\mcal{D} = \sum_i \mcal{D}_{\text{3D Space}}(i)$$
\end{enumerate}
\end{itshape}

Consider an arbitrary signal, $x(\r,t)$ which we decompose into orthonormal functions. The number of functions required to describe $x(\r,t)$ gives the dimensionality of the space.
\begin{align}\label{E:3d-decompose}
x(\r,t) &= \sum_i \beta_i \psi_i(\r,t)
\\
\psi_i(\r,t) &= 
\begin{cases}
	\Phi_i(\r,t;\k)
	&; \r\in\mbb{S},t\in[0,T]
	\\
	0 &;\ \text{else}
\end{cases}
\end{align}

$\Phi_i(\r,t;\k)$ is an eigenfunction of the Helmholtz equation~\cite[6.94.1]{Gradshteyn00} in 3-D cartesian coordinates:
\begin{align}\label{E:helmholtz-eig}
\Phi_i(\r,t;\k)&=\frac{1}{\sqrt{T}}\exp\left[\j\left(k_x x + k_y y + k_z z + ckt\right)\right]
\\
&=\varphi_i(\r;\k)\frac{1}{\sqrt{T}} \exp (\j c k t)
=\varphi_i(\r;\k) \phi_i(t)\label{E:helmholtz-eig-2}
\end{align}
where 
$\k$  is the vector wave-number, 
$\k=(k_x,k_y,k_z) 
$ 
and $k=\|\k\|$ is the magnitude, commonly called the scalar wave-number.
 %

In  Problem~\ref{Prob:2wt} the enumeration of orthonormal functions was explicit -- since there was a linear sum~\eqref{E:DFT}. In the case of 3D wave fields, each orthonormal time-bandwidth function $\phi_i(t)$ is now attached to several spatial-functions $\varphi(t)$: as shown in \eqref{E:helmholtz-eig-2}. The enumeration is carried out by choosing the wave vector to be an integer combination:
\begin{equation}
(k_x,k_y,k_z) = \frac{1}{R}\left(\eta_x,\eta_y,\eta_z\right), \ \eta_x,\eta_y,\eta_z\in\field{Z}
\end{equation}
We thus restrict the wave-number $k$ to be integer multiples of 
\begin{equation}\label{E:integer-k}
k=\frac{2\pi }{cT}i \quad i\in\field{Z}.
\end{equation}


%
 %
Using the Jacobi-Anger  expansion~\cite[8.524.1]{Gradshteyn00} write \eqref{E:helmholtz-eig-2}
\begin{equation}
\varphi(\r;\k)=4\pi\sum_{n=0}^\infty\j^n j_n(k|\r|)\sum_{m=-n}^n Y_n^m(\hat{\r})Y_n^m(\hat{\k})\label{E:rodversion}
\end{equation}
From~\cite{JonKenAbh:ICASSP02}, $\varphi(\r;\k)$ may be truncated at 
\begin{equation}
n<\mcal{N}_{(R,k)}=\left\lceil\frac{e k R}{2}\right\rceil
\end{equation}
which results in
\begin{equation}
\mcal{D}_{(R;k)}=\sum_{n=0}^{\mcal{N}_{(R,k)}}\sum_{m=-n}^n1=\left[\mcal{N}_{(R,k)}+1\right]^2
\end{equation}
terms for each value $k$.
The dimensionality is given,  (simlarly to Problem~\ref{Prob:2wt}) by counting the number of terms in the sum~\eqref{E:3d-decompose}. 
\begin{equation}\label{E:3d-super-sum}
x(\r,t)=\frac{1}{\sqrt{T}}\sum_i \beta_i \varphi_i(\r;\k) \exp\left(\frac{2\pi i}{T}t\right)
\end{equation}
As for Problem~\ref{Prob:2wt}, the value of $i$ in \eqref{E:integer-k} is limited by the frequency range of interest:
\begin{equation}
F_o-W\leq \frac{i}{T} \leq F_o+W
\end{equation}

\begin{align}
\mcal{D} &= \sum_{i=(F_o-W)T}^{(F_o+W)T} \mcal{D}_{R;k(i)} = \sum_{i=(F_o-W)T}^{(F_o+W)T} 
\left[\mcal{N}_{(R,k(i))}+1\right]^2
\\
&=\sum_{i=0}^{2WT} \left(\left\lceil\frac{e\pi R(F_o-W+i)}{cT}\right\rceil+1\right)^2
\label{E:3d-sum}
\end{align}
The first term in \eqref{E:3d-sum} is the integer $(N_0+1)^2$, since $i=0$. For any integer $i>0$, the ceiling causes an increment to $(N_0+2)^2$. There are $cT/(e\pi R)$ terms before the series contains $(N_0+3)^2$ (at the next increment)
\begin{equation*}
\mcal{D} = \overbrace{(N_0+1)^2+\underbrace{(N_0+2)^2+\cdots}_{cT/(e\pi R)\text{ terms}} + \underbrace{(N_0+3)^2+\cdots}_{cT/(e\pi R)\text{ terms}}+\cdots}^{2WT+1\text{ terms}}
\end{equation*}
with $cT/(e\pi R)$ terms in each set above. Re-write \eqref{E:3d-sum}, with a new index $j$ over sets (removing the ceiling function). In \eqref{E:3d-sum} there are $2WT+1$ terms, while there are $$2WT \frac{e\pi R}{cT}$$  sets, then 
\begin{equation}\label{E:3d-sequence-sum}
\mcal{D}\leq(N_0+1)^2+\frac{cT}{e\pi R}\sum_{j=1}^{2WT \frac{e\pi R}{cT}}(N_0+j+1)^2  
\end{equation}
Evaluating the summation \eqref{E:3d-sequence-sum} completes the proof.
\end{proof}

\begin{proof}[Theorem~\ref{thm:1}]
\begin{enumerate}
\item Write a time-space function in terms of $$\psi_i(\r,t) = \Psi_i\left(\r;k(i)\right) \exp\left[-\j  k(i) c t\right]$$
\item Observe that every spatial function $\Psi_i(\r,t)$ which satisfies \eqref{E:helmholtz} in two dimensions has $$\mcal{D}_{\text{2D Space}}(i) = e \pi F(i) R/c + 1$$ degrees of freedom.

\item The total degrees of freedom is given by a summation (equivalent to \eqref{E:xtsum})
$$\mcal{D} = \sum_i \mcal{D}_{\text{2D Space}}(i)$$
\item The same observation as for \eqref{E:3d-sequence-sum} is needed, and leads to the result.
\end{enumerate}
\end{proof}

\end{document}